\documentclass[preprint2]{aastex}
\usepackage{graphicx,latexsym,amsmath,amssymb}
\usepackage{natbib}
\usepackage{hyperref}
\usepackage{lastpage}

\def\simlt{\lower.5ex\hbox{$\; \buildrel < \over \sim \;$}}
\def\simgt{\lower.5ex\hbox{$\; \buildrel > \over \sim \;$}}

\bibpunct[,]{(}{)}{;}{a}{}{,}

\shorttitle{Tidal disruption events in circumbinary disks}
\shortauthors{Brem et al} 

\begin{document}

\title{Tidal disruptions in circumbinary disks. II: Observational signatures in the reverberation spectra}
\renewcommand{\thefootnote}{\alph{footnote}}

\author{P. Brem$^1$\footnote{Patrick.Brem@aei.mpg.de}  , J. Cuadra$^2$, P. Amaro-Seoane$^1$ \& S. Komossa$^{3}$}
\affil{$^1$Max Planck Institut f\"ur Gravitationsphysik (Albert-Einstein-Institut), D-14476 Potsdam, Germany}
\affil{$^2$Instituto de Astrof\'isica, Facultad de F\'isica, Pontificia Universidad Cat\'olica de Chile, Santiago, Chile}
\affil{$^3$Max Planck Institut f\"ur Radioastronomie, Auf dem Huegel 69, 53121, Bonn, Germany }

\begin{abstract}
%
%
Supermassive Binary Black Holes (SMBBHs) with sub-pc separations form in the course of galaxy mergers, if both galaxies harbour massive black holes. Clear observational evidence for them however still eludes us.
%
%
We propose a novel method of identifying these systems by means of reverberation mapping their circumbinary disk after a tidal disruption event has ionized it. The tidal disruption of a star at the secondary leads to strong asymmetries in the disk response.
%
%
We model the shape of the velocity--delay maps for various toy disk models and more realistic gas distributions obtained by SPH simulations. The emissivity of the ionized disk is calculated with {\em Cloudy}.
%
%
We find peculiar asymmetries in the maps for off center ionizing sources that may help us constrain geometrical parameters of a circumbinary disk such as semi-major axis and orbital phase of the secondary, as well as help strengthen the observational evidence for sub-parsec SMBBHs as such.
%
%

\end{abstract}


\section{Introduction}

The theory of hierarchical structure formation predicts that mergers between galaxies are common in our universe. If we believe that most galaxies host a central massive black hole, then dynamical friction will bring those to the center of the newly formed galaxy and they will inevitably form a SMBBH. Interaction with stars and gas will then shrink the binary to sub-parsec separations \citep[e.g.,][]{BegelmanEtAl80, ColpiDotti11}.

Observational evidence exists for the stage previous to the formation of binaries, when the two black holes are still sinking into the galaxy and appear as dual active galactic nuclei (AGN; \cite{KomossaEtAl03}). However, the closest projected separation observed for such a system is still 7 pc \citep{RodriguezEtAl06}. On smaller scales, resolving the two individual AGN spatially is not possible and we must rely on indirect evidence to identify SMBBH {\it candidates}.  Such is the case of the blazar OJ 287. Its almost-periodic luminosity fluctuations
      can be explained by the presence of a sub-parsec separation SMBBH. Different
      scenarios and geometries of the system have been proposed and explored
      (e.g., \cite{SillanpaaEtAl88}). 
\citeauthor{ValtonenEtAl08} (e.g., \citeyear{ValtonenEtAl08})
      favor a scenario where the secondary crosses the accretion disk of the primary twice per orbit. 

In this work, we propose a novel method for identifying SMBBHs, based on the reverberation mapping technique.
For standard AGN, broad emission lines from a variety of elements are observed. They originate in cold gas clouds on Keplerian orbits around the accreting black hole, which are  ionized by the AGN emission and emit recombination lines. 

For the BLR in AGN, reverberation mapping has been applied to measure both the distance of the clouds to the ionizing source as well as its projected velocity \citep{BlandfordMcKee82,Peterson93,GrierEtAl13}. The distance is obtained by extracting the time-delay between variations in the continuum flux and the broad line flux.
The projected velocity is given directly by the Doppler shift. 

If we consider a SMBBH with separation below one parsec, accretion is mostly suppressed because the binary orbital motion clears out an inner cavity in the accretion disk, a region of very low density \citep[e.g.,][]{ArtymowiczLubow94, delValleEscala12}.
In \citep{AmaroEtAl2012} (hereafter, Paper I) we have modelled the evolution of such a system under the assumption that the disk fragments and forms stars.  
We found that some of the stars get into orbits that approach very close to either MBH and are tidally disrupted. The event rate for these tidal disruption events (TDEs) we find is of the order of $10^{-5} - 10^{-4} {\rm yr}^{-1}$ per system. An independent, semi-analytical study by \cite{LiuChen13} shows that
more than $20 \%$ of all observable TDEs could happen in SMBBHs and a substantial fraction of those are expected to be surrounded by a circumbinary gas disk.

After stellar disruption, a fraction of the stellar material is accreted by the SMBH, producing a luminous flare of radiation (e.g., \cite{Rees88}).
   About two dozen TDEs and candidates have been identified by now (review by \cite{Komossa2012}),
   including in X-rays \citep{KomossaBade99}, the UV \citep{GezariEtAl06},
   $\gamma$-rays \citep{BloomEtAl11,BurrowsEtAl11}, and the radio \citep{ZaudererEtAl11}.
   If the TDE occurs in a gas-rich environment, its continuum emission is reprocessed into
   emission lines. Several candidate events with transient Balmer lines, HeII and high-ionization
   iron lines have been identified in recent years \citep{KomossaEtAl08apjl}.
   In case of an SMBBH, the flare will also ionize the circumbinary disk.  With the same techniques used for reverberation mapping of BLRs in AGN, we investigate the peculiar shape of the velocity--delay maps that can be obtained from observing the afterglow of these events. In particular, if the TDE happens at the secondary, it might be offset from the center by as much as half the cavity radius. This leads to a tilt in the velocity-delay map, which depends not only on the  physical parameters of the system, but also on the orbital phase of the SMBBH at the time of the TDE. The observation of this tilt will then not only allow us to identify a SMBBH candidate, but also to characterise it.

\section{Method}
\label{sec.method}

\subsection{Physical picture}

We consider here a SMBBH that is the product of a recent, sufficiently gas rich, galaxy merger. Once dynamical friction has brought the two MBHs to the center of the newly formed galaxy, three body scattering of stars and the interaction with the surrounding gas will harden the binary further, leading to the sub-parsec separations that are of interest here (see \cite{ColpiDotti11}, for a review). The alignment of MBH orbital angular momentum and the surrounding gas leads to the formation of a circumbinary disk (e.g., \cite{MayerEtAl07,DottiEtAl09}). In this evolutionary stage, the torque of the orbital MBH motion clears out a cavity in that disk \citep{delValleEscala12}. Due to the cavity, the continuous gas inflow onto either MBH can be very small and we do not expect the SMBHs to emit a significant amount of AGN-like radiation (e.g., \cite{MacFadyenMilosavljevic08}) . This is why we expect a possible TDE to dramatically increase the irradiation onto the disk.

\subsection{Reverberation mapping}

The continuum photons emitted by the TDE can be absorbed by gas and trigger recombination, leading to the emission of an ``echo'' from the gas. Formally, we are looking for the transfer function $\Psi_l(v,t)$ \citep{BlandfordMcKee82}, defined as

\begin{equation}
  L_l(v,t) = \int_{-\infty}^{\infty} dt' L_c(t') \Psi_l(v,t-t').
\label{eq.trans}
\end{equation}
$L_l(v,t)$ is the recombination luminosity in a certain line at time $t$ and doppler shifted corresponding to a velocity $v$. $L_c$ describes the total, time dependent continuum luminosity emitted by the TDE flare. For $L_c(t) \propto \delta(t)$, 
$\Psi_l$ is simply equal to the response $L_l$. For the purpose of deriving mock transfer functions (also called {\em velocity delay maps}), we can directly compute $\Psi_l$ using this equality. With real AGN data, however, the inverse problem of deriving $\Psi_l$ given some noisy $L_c(t)$ and $L_l(t)$ is very challenging (see e.g. \cite{PetersonEtAl04} for a review).

Given a gas distribution and an observer at $(0,0,z_0)$, with $z_0 \gg$ the size of the system, the luminosity response for a volume-element $dV$ at position $\vec r = (x,y,z)$ moving with velocity $v$ at time $t$ is

\begin{equation}
  dL_l(v,t) = j_l(\vec r,t-(z_0-z)/c) dV,
\label{eq:lum}
\end{equation}

considering that the light had to travel a distance $z_0-z$ from the volume element to the observer, and the emissivity $j_l$ itself is a delayed response to the continuum luminosity $L_c$ from the source,

\begin{equation}
  j_l(\vec r,t) \propto L_c(t-r/c).
\end{equation}

We thus have two retardation effects to be taken into account in the computation of the transfer function. The velocity-delay maps presented here are obtained by integrating Eq. \ref{eq:lum} over the whole disk.

\subsection{Computing the emissivity}

We aim to produce mock velocity-delay maps of the response of the gas in a circumbinary disk around a SMBBH after a TDE.  For this we need to assume a geometry for the gas distribution and a spectrum of the ionizing continuum radiation from the TDE. Once this is defined, we use the radiation transfer code {\em Cloudy} to obtain the emissivity for certain recombination lines as a function of the location. Calculations were performed with version 13.02 of {\em Cloudy}, last described by \cite{Ferland13}. This gives us the contribution in the velocity-delay plane.

First, we model the TDE spectrum as a black-body with an effective temperature \citep{Ulmer99} given by
\begin{equation}
T_{\rm eff} = 2.5 \times10^5 M_6^{1/12} R_{\star,\odot}^{-1/2} M_{\star,\odot}^{-1/6} \,K,
\label{eq.temp}
\end{equation}

where $M_6 = M_\bullet/10^6 M_\odot$ is the mass of the MBH disrupting the star in units of $10^6 M_\odot$, $R_{\star,\odot}$ is the radius of the disrupted star in solar radii and $M_{\star,\odot}$ is its mass in solar masses.
Considering a solar-type star and $M_6 \sim 1$, we fix the fiducial value at $T_{\rm eff} = 3 \times 10^5 K$. Assuming the emitting region is the surface area $A$ of the accretion disk up to a radius $R$ we find a total peak continuum luminosity of

\begin{align}
  L &= \sigma A T_{\rm eff}^4 = 2 \pi R^2 \sigma T_{\rm eff}^4\nonumber \\
    &= 10^{44} {\rm erg}\,{\rm s}^{-1} \left( \frac{R}{40 r_g}\right)^2 \left( \frac{T_{\rm eff}}{3\times 10^5 K} \right)^4,
\label{eq.lum}
\end{align} 
which means we reach typical values for the (luminous) TDEs observed in X-rays (review by \cite{Komossa2012}) for an emitting region of $R = 40 r_g$ which lies just inside the tidal radius for a $10^6 M_\odot$ SMBH and a solar mass star. However, as we discuss later, lower values of $L$ and $T$ simply decrease the penetration depth of the continuum radiation into the disk and thus lead to an even sharper velocity delay map.

For the gas distribution we first assume a disk of constant hydrogen number density and solar metalicity. This disk is hollowed out around the SMBBH center of mass with a cavity radius of twice the binary semi-major axis, as expected for circumbinary disks of near-equal mass binaries. The disk is assumed geometrically thin, with a scale height of $h/r < 0.1$. Inspired by our numerical study in paper I, we adopt for simplicity the constant number density $n = 10^8 {\rm cm}^{-3}$ and a SMBBH semi-major axis of $a = 0.01\,{\rm pc}$.

\begin{figure}
\plotone{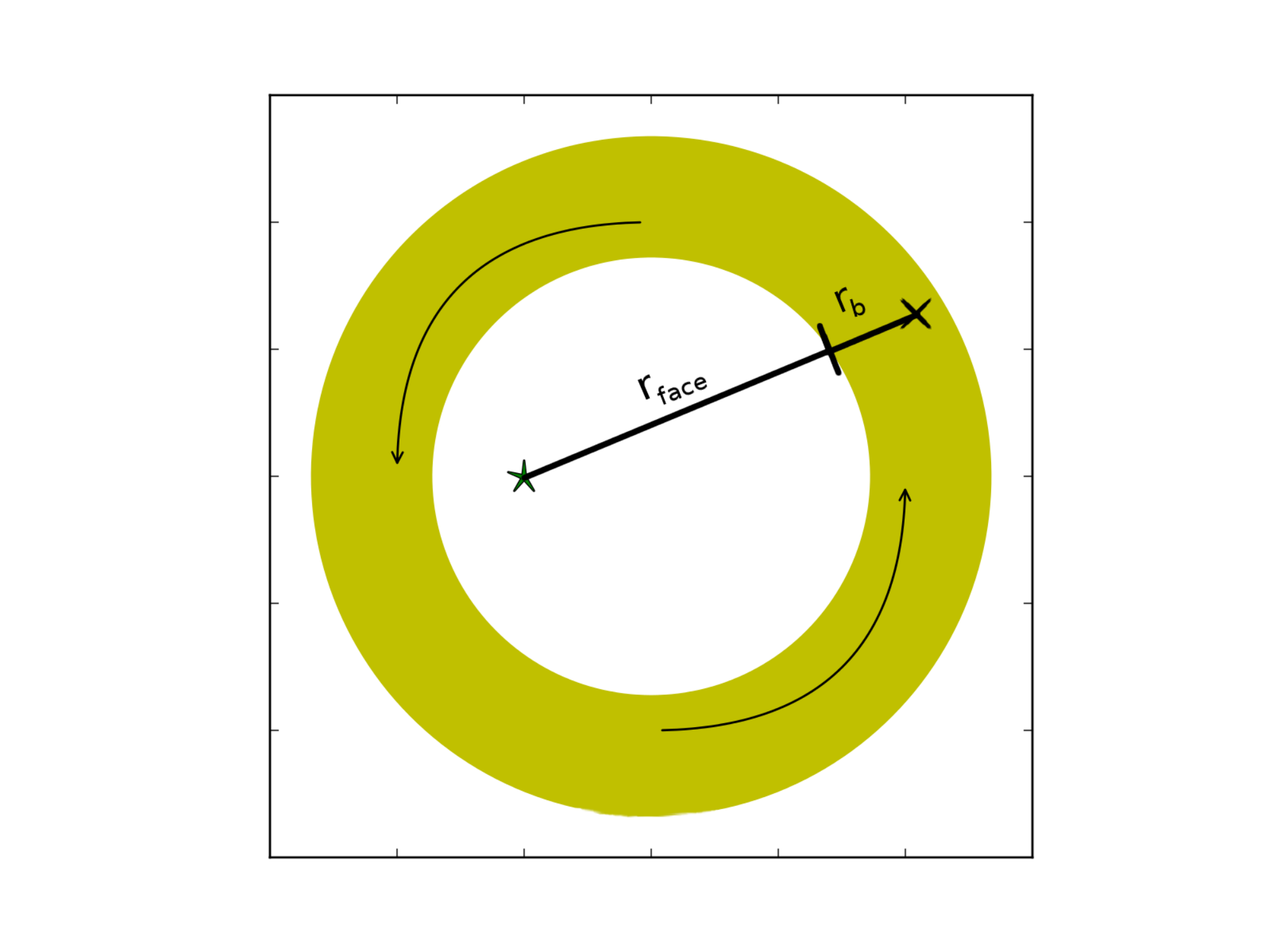}
\caption
   {Illustration of $r_{\rm face}$ and $r_b$ that are used to interpolate the emissivity profiles using {\em Cloudy}. The continuum source is located at the star inside the gap.
}
\label{fig:sketch}
\end{figure}

The emissivity for each line per unit volume depends on the distance of the face of the gas disk to the continuum source, $r_{\rm face}$ and the distance of the volume element behind the face, $r_b = r - r_{\rm face}$ for all $r > r_{\rm face}$. Figure \ref{fig:sketch} shows a schematic of the different radii used as an input for {\em Cloudy}.

For tidal disruptions at the secondary, the center of the continuum source can be significantly offset from the center of mass of the binary. Thus, we need to obtain the emissivity profiles for a variety of source-face distances $r_{\rm face}$. This is achieved by varying the inner cloud radius in {\em Cloudy} from the smallest ($r_{\rm face} = a$) to the largest ($r_{\rm face} = 3a$) possible value. We  then interpolate on this grid for each $r_{\rm face}$ we require and get the emissivity profile $j_l(r_b)$ for each volume element $dV$ even with an offset source. The line luminosity of this volume element is then $dL_l = j_l(r_b) dV$.
 
For small number densities such as our fiducial value of $n = 10^8 {\rm cm}^{-3}$, different emission lines have significantly different emissivities at different $r_b$ and this spread is important. For larger densities $n > 10^{10} {\rm cm}^{-3}$, however, the emission drops off quickly within $r_b \ll r_{\rm face}$. Thus we can neglect retardation effects within the disk and safely assume that all the emission comes from the very inner edge of the disk.

After obtaining the emissivity profiles, we perform the geometrical operations of rotating the system into a specific inclination angle and obtaining the delay for each volume element of the disk. Together with its respective line emissivity, we can construct the velocity-delay map for arbitrary source positions and inclinations.

\subsection{Mock spectra}

Measuring actual transfer functions requires dense spectroscopic monitoring right after a TDE and is therefore a very challenging task. We thus want to produce mock spectra for several points in time as well, for which we need to adopt an explicit form for $L_c(t)$. Then the observed spectrum, proportional to $L_l(v,t)$, can be computed by explicitly integrating Eq. \ref{eq.trans}. We adopt for the continuum luminosity

\begin{equation}
  L_c(t) \propto \begin{cases}
    0 & t < t_{\rm min} \\
  \left( \frac{t}{t_{\rm min}} \right)^{-5/3} & t \ge t_{\rm min} \\
  \end{cases}
,
\end{equation}

where $t_{\rm min}$ is the time when the first bound material reaches the MBH after disruption, which is given by \citep{Ulmer99}

\begin{equation}
  t_{\rm min} = 0.11 \, R_{\star,\odot}^{3/2} \, M_{\star,\odot}^{-1} \, M_6^{1/2} \, {\rm yr}.
\end{equation}

We adopt the fiducial value $t_{\rm min} = 0.11 \, {\rm yr}$.

\section{Results}
\label{sec.results}

\begin{figure*}
\begin{center}
          {\includegraphics[clip=true,trim=0 0 0 1cm, width=\textwidth,height=\textheight]{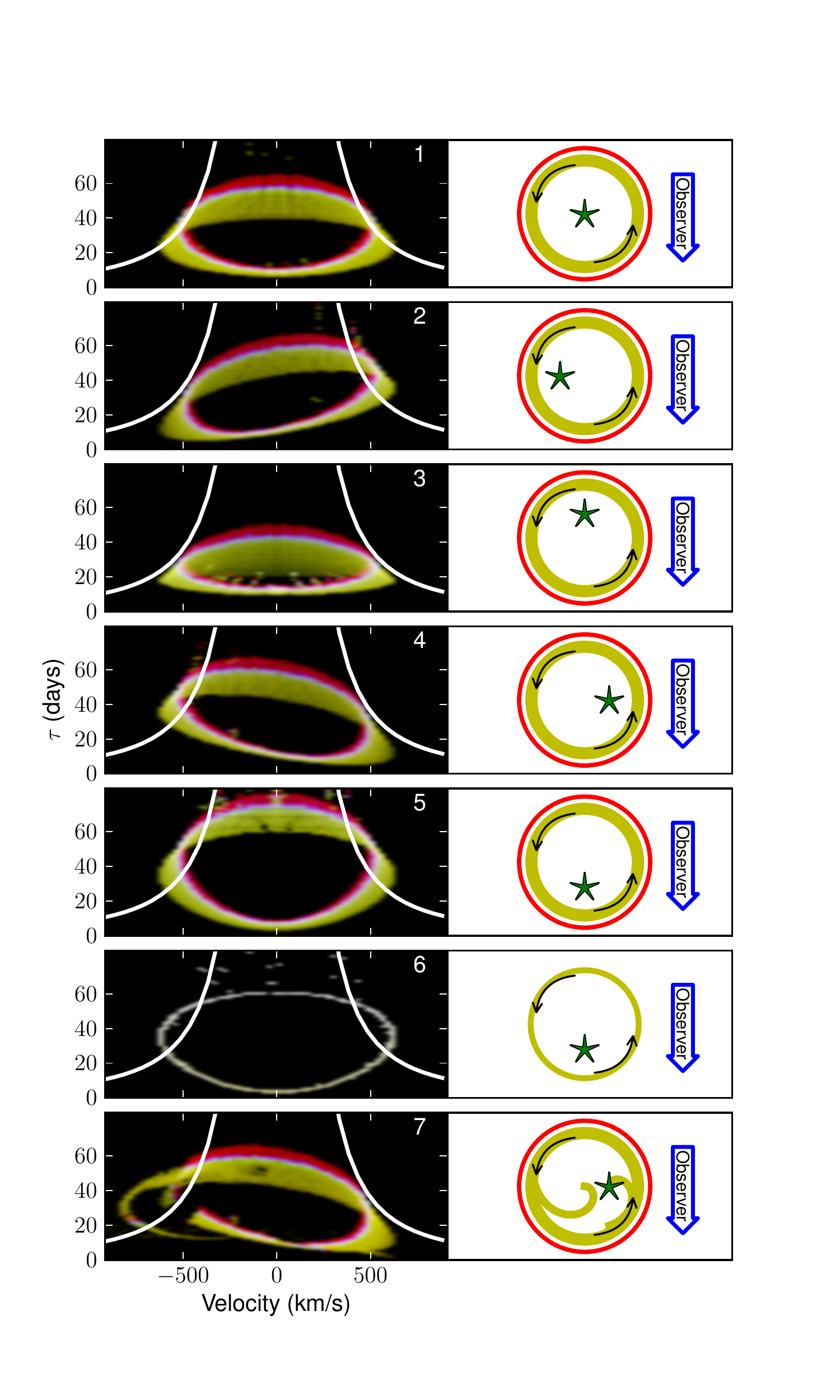}}
\caption
   {
Velocity-delay maps for a perfectly circular gas disk with a cavity size of $r_{\rm cavity} = 0.02\,{\rm pc}$. All maps are computed for an observational inclination of 45 degrees. The right column shows a schematic of the position of the tidal disruption and the direction to the observer. The white lines show the ``virial envelope'', given by $v^2 = G M_\bullet/ (c \tau)$.
   }
\label{fig:delayall}
\end{center}
\end{figure*}

\begin{figure}
\plotone{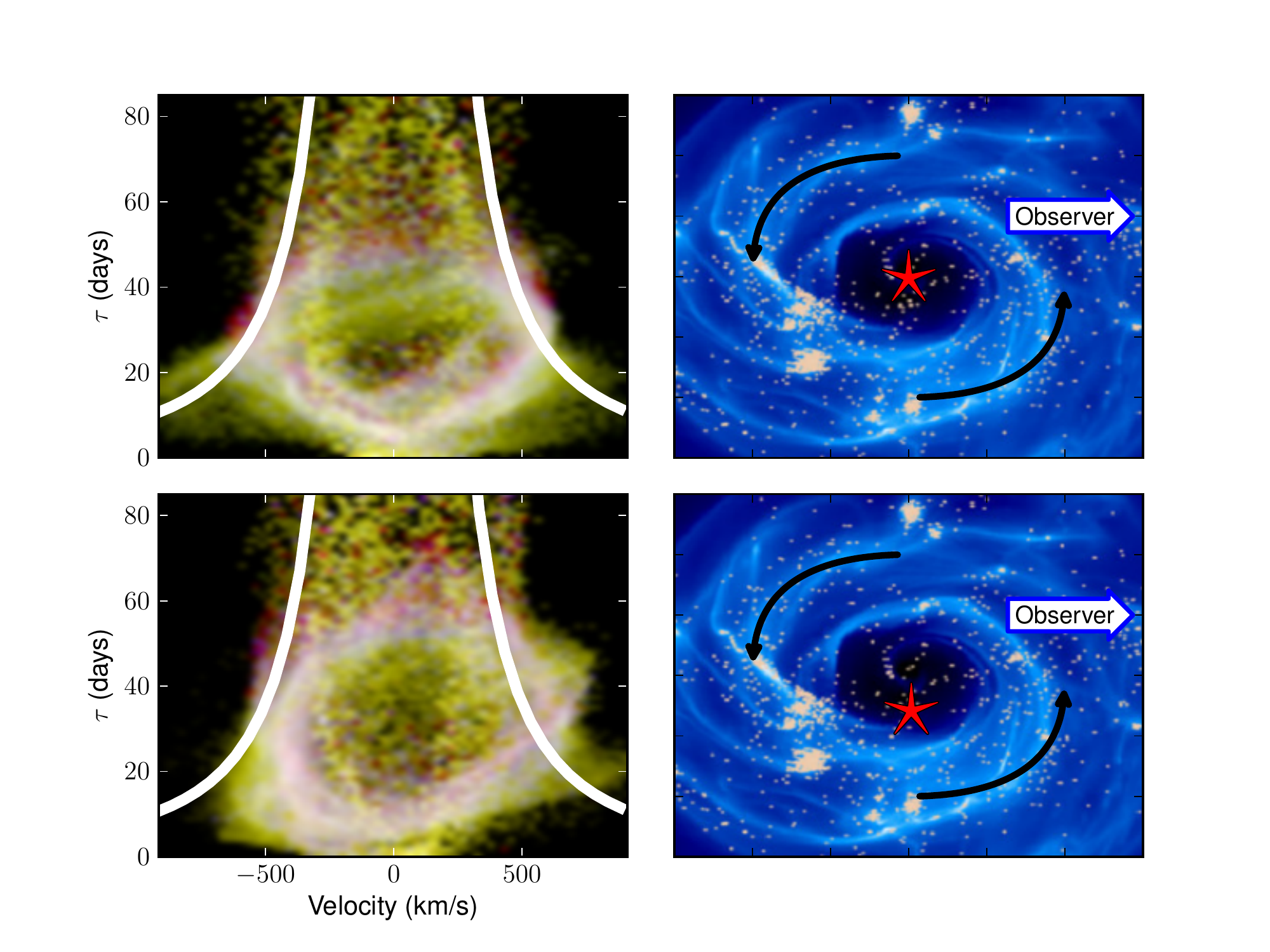}
\caption
   {
Velocity-delay map for a gas distribution taken from one of our \mbox{GADGET} simulations, RGB encoded as Fig.~\ref{fig:delayall}. An off-centered continuum source leads to a tilt in the response curve. The white contribution marks the region where all three shown lines contribute significantly. Notice that the direction to the observer is different than in Fig.~\ref{fig:delayall}.
   }
\label{fig:delaybeta}
\end{figure}

\begin{figure}
  \plotone{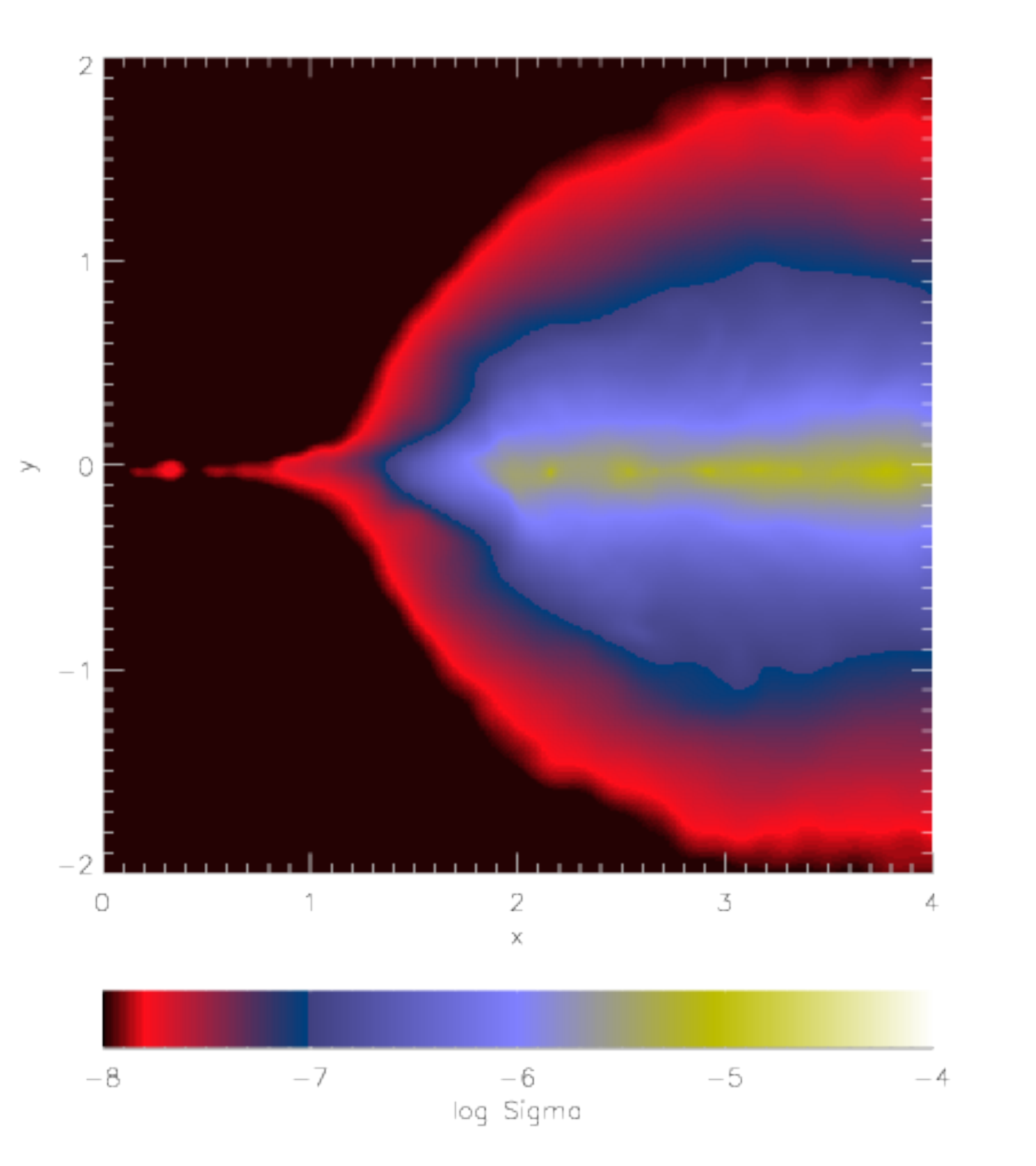}
\caption{Cross section of the surface density of the disk used in Fig.~\ref{fig:delaybeta}, integrated along the azimuthal angle.}
\label{fig:cross}
\end{figure}

Fig. \ref{fig:delayall} shows the velocity-delay maps for a TDE flare in seven different physical situations.
In the left panels, the RGB images show the emission in three different emission lines: H$\beta$ is given the red component, HeII$\lambda 4686$ the green and HeI$\lambda 5876$ the blue. With our assumptions for the effective temperature and luminosity, H$\beta$ emission extends from the inner to the outer edge of the emitting region, while HeII and HeI drop off at a certain depth. The dominant yellow colour traces the region where both H$\beta$ and HeII are strong. The white ring marks the intermediate region where all three lines show high contributions, while the red part tracks the outer regions where the Helium lines have vanished. 

The first row, case 1, shows the map produced by a TDE occuring at the centre of the disk, which is the case expected for a single MBH.  This map can be directly compared with Fig.~14 (c) in \cite{GrierEtAl13}.  In this case the map is completely symmetric with respect to the $v=0$ axis.  

The following rows, cases 2--5, show the situation for a significantly off-centre TDE, which is what we expect if the TDE happens at the secondary of an unequal mass SMBBH. In these examples the TDE is offset by $a$ with respect to the center, which is the typical offset for a lower mass secondary.
We want to highlight the most significant features of the velocity-maps as compared to the picture for axisymmetric distributions of BLR clouds. First, in the orbital phases where the secondary is close to either the red- or blue-shifted edge of the disk, case 2 and 4, we see the first response tilted to the red or blue, respectively. This asymmetry corresponds to the continuum emission not being at the center of the disk.

The second feature, most prominently in the orbital phases close to the front of the disk, case 5, as seen from the observer, is the large void in the middle of the map. Because the gas directed towards us receives the continuum after a time of only $a/c$ and the back part after $3a/c$, there is a large delay in the non-shifted parts of the emission lines. This void can also be smaller than expected for a central continuum source if the TDE happens close to the back part of the disk, case 3, but only appears if the geometry resembles a planar disk. In a spherical distribution of gas, we see an echo at zero velocities at all times instead.

The third feature, which can also be seen for regular BLR observations, is the delay for the first emission, which here also depends on the phase of the binary orbit.

In case 6, we show the map for a thicker disk of density $n = 10^{11} {\rm cm}^{-3}$, for which all of the ionizing radiation is deposited in the very surface of the disk and all emission lines originate from the same region. The dots outside the circle are artifacts from the discrete angular grid we use for ray-tracing.

In the last row we show the map for accretion streams falling onto both black holes from the inner edge of the disk \citep[see, e.g.,][]{MacFadyenMilosavljevic08, CuadraEtAl09}. Their features are apparent as higher velocity contributions to the left side of the original response map.

Fig.~\ref{fig:delaybeta} shows the simulated velocity-delay map for a gas distribution taken from one of our \mbox{GADGET} simulations of Paper I. The top panel shows the reverberation for a central continuum source, the bottom panel for the ionizing source offset from the center. The structures are less well-defined than for the idealised circular disk, however we still see a clear tilt to the negative velocities. The time difference between the peak in the most red- and blue-shifted echo is about 10 days. The binary orbital phase corresponds to case 2 of the perfect disk model explained above. White dots in the figure are newly formed stars which are not taken into account when modelling the reprocessed line profiles. A cross section of the disk is shown in Fig.~\ref{fig:cross}.

\begin{figure}
{\includegraphics[clip=true, trim=0.75cm 0cm 0cm 0cm, width=\columnwidth]{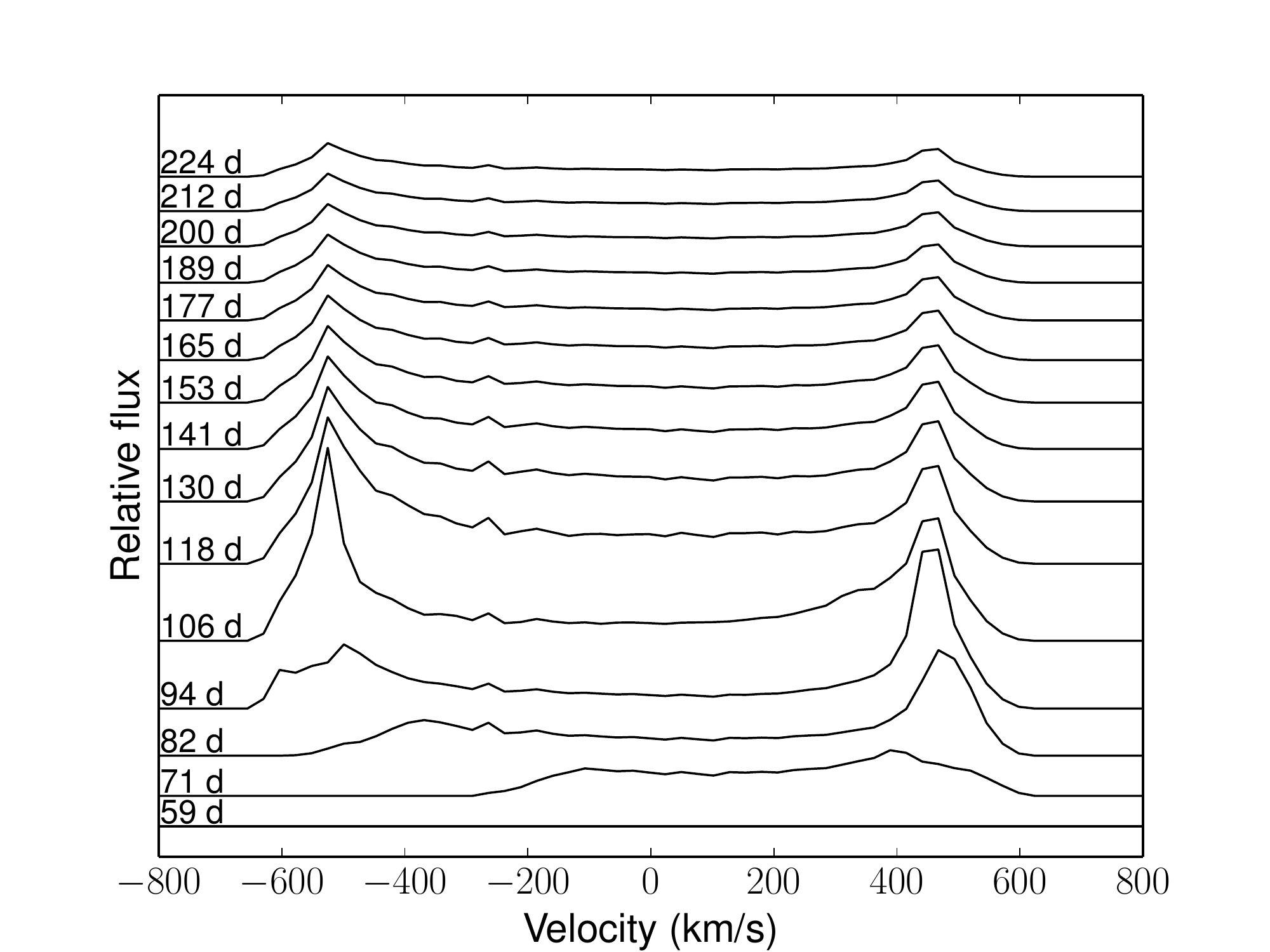}}
\caption{Mock spectra for $H_\beta$ of case 4 of Fig.~\ref{fig:delayall} }
\label{fig.spectra}
\end{figure}

In Fig.~\ref{fig.spectra} we show the mock spectra that case 4 of Fig.~\ref{fig:delayall} would produce. The blue-shifted part of the broadened line is rising first, but after about 100 days the red-shifted part starts to outshine it. Even without obtaining the two dimensional velocity delay map this shift is very prominent in the line spectrum alone, allowing one to observe the SMBBH signature even without the dense spectroscopic campaigns necessary for reverberation mapping.

\section{Discussion}
\label{sec.discussion}

We showed velocity-delay maps for 3 emission lines in a perfectly circular, circumbinary disk and a more realistic snapshot from one of our \mbox{GADGET} simulations. Both show peculiar features that distinguish them clearly from a typical spherical or axisymmetric BLR. While the exact shape of the maps will depend on the physical properties of the system, most prominently the TDE spectrum and luminosity and the gas density and composition, the general features such as the void and the tilt will remain even for very different conditions.

\subsection{Assumptions about the ionizing continuum and gas}

Throughout this work we have assumed the continuum source to be a TDE flare, with the effective temperature and total luminosity given by Eq. \ref{eq.temp} and \ref{eq.lum}. However, we do not need these specific assumptions to hold in order to observe similar velocity delay maps. While the spectral energy distribution and total luminosity of the ionizing source will change the penetration depth and thus the extent of the reprocessing regions for each element, the transfer function will not change qualitatively if we assume, for example, a standard AGN mini-disk inside the cavity as the source for the radiation.

We note that compared to the complex variability in AGN, which makes it rather complicated to obtain the transfer function of the line response, the TDE flux rises on a timescale of days, which may give far better time resolution in obtaining the velocity-delay map.

For gas densities as low as $n = 10^8 {\rm cm}^{-3}$ the emitting regions for different ionization lines are nicely separated. If we raise the gas density in the disk, the penetration length of the ionizing continuum will become shorter and thus the rings will be localized to the inner edge of the disk, as shown in the bottom panel of Fig. \ref{fig:delayall}. A colder effective temperature and lower luminosity of the TDE flare will have the same effect.

The temperature profile of an accretion disk around a single SMBH can be approximated under the assumption that the accreted mass is converted into energy and radiated as blackbody radiation 

\begin{equation}
  \sigma T^4(r) = \frac{3}{8 \pi} \left(1-\sqrt{\frac{R_{\rm ISCO}}{r}}\right)\dot M \Omega_K^2,
\end{equation}
where $\sigma$ is the Stefan-Boltzmann constant, $R_{\rm ISCO} = 6 G M_\bullet/c^2$, $\Omega_K$ the Keplerian angular velocity and $\dot M$ the mass accretion rate. This leads to the following temperature profile for $R \gg R_{\rm ISCO}$:

\begin{equation}
  T(r) = 180 {\rm K} \left( \frac{r}{0.01 {\rm pc}}\right)^{-3/4} \left( \frac{M}{10^6 M_\odot}\right)^{1/4},
\end{equation}
where we have assumed the accretion rate to be $\dot M = \dot m \frac{L_{\rm Edd}}{\epsilon c^2}$ with both radiative efficiency $\epsilon = 0.1$ and $\dot m = 0.1$. This shows that at the inner edge of the cavity at $r \approx 0.01 {\rm pc}$ the disk is expected to be cold, and all the hydrogen can safely be assumed to be in its ground state before the TDE occurs.

\subsection{Other observable signatures and future observations}

Beyond the reverberation signal itself, there is a number of other
  observational consequences of the events discussed here.
  Firstly, we predicted the total luminosity of H$\beta$ from the disk. If we assume a fully covering sphere of gas then {\em Cloudy} tells us that for the model tidal disruption spectrum we consider here, the total luminosity in H$\beta$ will be $L_{{\rm H}\beta,{\rm full}} \approx 10^{41} {\rm erg}\,{\rm s}^{-1}$ for an ionizing luminosity of $L = 5 \times 10^{44} {\rm erg}\,{\rm s}^{-1}$. Taking the disk to have a scale height of about $h/r = 0.1$, this results in a covering fraction of $f = 0.05$ and thus an upper limit for the reprocessed line luminosity of $L_{{\rm H}\beta,{\rm disk}} = 5 \times 10^{39} {\rm erg}\,{\rm s}^{-1}$. As we have seen from SPH simulations, the disk can be significantly thicker than that at the inner rim and thus produce larger luminosities.
We note in passing that there are additional sources of line emission, not considered here. These would look very different from the emission-line signature discussed here. First, single stellar streams, bound and/or unbound, would reprocess some of the radiation themselves (e.g., \cite{StrubbeQuataert09}). However, their kinematic signature would be very different from a circum-binary disk. Second, distant interstellar matter will reprocess some of the radiation into emission lines (e.g., \cite{KomossaEtAl08apjl}). These  lines will be narrow, and consist mostly of forbidden transitions, and they will reflect the much lower density of the ISM.

Second, we note that the properties of the circumbinary
disk we discussed here are not too different from the BLR in AGN
(with BLR densities around $10^{9-12} {\rm cm}^{-3}$, within the range
considered here; e.g., \cite{OsterbrockFerland06,SneddenGaskell99}).
Therefore, if a classical BLR is present in these systems,
our gross predictions for delays also hold for a BLR. In particular, a BLR
would typically be more extended, and perhaps
more spherical, than the disk parameters considered
here, and so would add an extra signal at long
time delays (e.g., \cite{Peterson93,Peterson12,GrierEtAl13,BarthEtAl13,PancoastEtAl14}. Notice that this is different from other
methods for finding SMBBH candidates that assume the existence of
separate BLRs around either black hole (e.g, \cite{ShenLoeb10})

Third, \cite{Tanaka13} suggests an alternative continuum source which shows similar peak luminosities and effective temperatures as TDEs: Periodic gas infall from the edge of the circum-binary disk onto one of the binary companions. While this phenomenon, too, could produce a reverberation signal, it would look different from TDEs in several ways: TDEs are unlikely to repeat within decades or centuries, while stream-feeding would re-occur on timescales of weeks to years depending on the orbital parameters. Moreover, the recurrence of the accretion periods would cause some semi-permanent line emission in any more distant and/or lower density gas, while these are absent prior to a TDE.

In summary, we have presented a method to compute velocity-delay maps given an arbitrary gas distribution and an ionising spectrum.  We applied this method to the case of TDEs happening inside, but off-centre, circumbinary disks,  and found features that present a new    observational signature of sub-parsec scale binaries.
    Upcoming all-sky and transient surveys will detect TDEs in the thousands,
    including the LSST in the optical \citep{IvezicEtAl08}, planned radio surveys like
    with the SKA \citep{MurphyEtAl12}, and the proposed soft X-ray transient mission
    Einstein Probe \citep{YuanEtAl13}.
    According to \cite{LiuChen13}, up to $20\%$
    of these events may occur in SMBBH systems. The first candidate SMBBH in a
    TDE has recently been identified based on characteristic features in its
    X-ray decline lightcurve \citep{LiuEtAl14}. Follow-up optical spectroscopy
    of events detected in ongoing and upcoming transient surveys will
    identify the TDEs in gas-rich environments. The characteristic reverberation
    signature predicted in this work then provides us with
    a new method of measuring the properties of SMBBHs and their gaseous environment
    in an important phase of their evolution, and in a regime which is not directly
    accessible by spatially resolved observations.

\section*{Acknowledgments}

We thank Patricia Ar\'evalo for illuminating discussions on
reverberation mapping.  JC acknowledges support from CONICYT-Chile
through FONDECYT (11100240), Basal (PFB0609), Anillo (ACT1101)
and Redes (12-0021) grants; and from the European Commission's Framework Programme 7,
through the Marie Curie International Research Staff Exchange Scheme
LACEGAL (PIRSES-GA-2010-269264).  JC and SK thank the organisers of
the Max Planck--Chile Research Seminar, held in Berlin in June 2012,
where this work started.
PA, JC, and SK would like to thank the Kavli Institute for Theoretical Physics (KITP)
for their hospitality and support during the program on ``A universe of black holes''.
This research was supported by the Transregio 7
``Gravitational Wave Astronomy'' financed by the Deutsche
Forschungsgemeinschaft DFG (German Research Foundation) and in part by the National Science Foundation under
Grant No. NSF PHY11-25915.


\begin{thebibliography}{38}
\expandafter\ifx\csname natexlab\endcsname\relax\def\natexlab#1{#1}\fi

\bibitem[{{Amaro-Seoane} {et~al.}(2013){Amaro-Seoane}, {Brem} \&
  {Cuadra}}]{AmaroEtAl2012}
{Amaro-Seoane} P., {Brem} P., {Cuadra} J., 2013, ApJ, 764, 14

\bibitem[{{Artymowicz} \& {Lubow}(1994)}]{ArtymowiczLubow94}
{Artymowicz} P., {Lubow} S.~H., 1994, ApJ, 421, 651

\bibitem[{{Barth} {et~al.}(2013){Barth}, {Pancoast}, {Bennert}, {Brewer},
  {Canalizo}, {Filippenko}, {Gates}, {Greene}, {Li}, {Malkan}, {Sand}, {Stern},
  {Treu}, {Woo}, {Assef}, {Bae}, {Buehler}, {Cenko}, {Clubb}, {Cooper},
  {Diamond-Stanic}, {H{\"o}nig}, {Joner}, {Laney}, {Lazarova}, {Nierenberg},
  {Silverman}, {Tollerud} \& {Walsh}}]{BarthEtAl13}
{Barth} A.~J., {et~al.}, 2013, ApJ, 769, 128

\bibitem[{{Begelman} {et~al.}(1980){Begelman}, {Blandford} \&
  {Rees}}]{BegelmanEtAl80}
{Begelman} M.~C., {Blandford} R.~D., {Rees} M.~J., 1980, Nat, 287, 307

\bibitem[{{Blandford} \& {McKee}(1982)}]{BlandfordMcKee82}
{Blandford} R.~D., {McKee} C.~F., 1982, ApJ, 255, 419

\bibitem[{{Bloom} {et~al.}(2011){Bloom}, {Giannios}, {Metzger}, {Cenko},
  {Perley}, {Butler}, {Tanvir}, {Levan}, {O'Brien}, {Strubbe}, {De Colle},
  {Ramirez-Ruiz}, {Lee}, {Nayakshin}, {Quataert}, {King}, {Cucchiara},
  {Guillochon}, {Bower}, {Fruchter}, {Morgan} \& {van der Horst}}]{BloomEtAl11}
{Bloom} J.~S., {et~al.}, 2011, Sci, 333, 203

\bibitem[{{Burrows} {et~al.}(2011){Burrows}, {Kennea}, {Ghisellini}, {Mangano},
  {Zhang}, {Page}, {Eracleous}, {Romano}, {Sakamoto}, {Falcone}, {Osborne},
  {Campana}, {Beardmore}, {Breeveld}, {Chester}, {Corbet}, {Covino},
  {Cummings}, {D'Avanzo}, {D'Elia}, {Esposito}, {Evans}, {Fugazza}, {Gelbord},
  {Hiroi}, {Holland}, {Huang}, {Im}, {Israel}, {Jeon}, {Jeon}, {Jun}, {Kawai},
  {Kim}, {Krimm}, {Marshall}, {P.~M{\'e}sz{\'a}ros}, {Negoro}, {Omodei},
  {Park}, {Perkins}, {Sugizaki}, {Sung}, {Tagliaferri}, {Troja}, {Ueda},
  {Urata}, {Usui}, {Antonelli}, {Barthelmy}, {Cusumano}, {Giommi}, {Melandri},
  {Perri}, {Racusin}, {Sbarufatti}, {Siegel} \& {Gehrels}}]{BurrowsEtAl11}
{Burrows} D.~N., {et~al.}, 2011, Nat, 476, 421

\bibitem[{{Colpi} \& {Dotti}(2011)}]{ColpiDotti11}
{Colpi} M., {Dotti} M., 2011, Advanced Science Letters, 4, 181

\bibitem[{{Cuadra} {et~al.}(2009){Cuadra}, {Armitage}, {Alexander} \&
  {Begelman}}]{CuadraEtAl09}
{Cuadra} J., {Armitage} P.~J., {Alexander} R.~D., {Begelman} M.~C., 2009,
  MNRAS, 393, 1423

\bibitem[{{del Valle} \& {Escala}(2012)}]{delValleEscala12}
{del Valle} L., {Escala} A., 2012, ApJ, 761, 31

\bibitem[{{Dotti} {et~al.}(2009){Dotti}, {Ruszkowski}, {Paredi}, {Colpi},
  {Volonteri} \& {Haardt}}]{DottiEtAl09}
{Dotti} M., {Ruszkowski} M., {Paredi} L., {Colpi} M., {Volonteri} M., {Haardt}
  F., 2009, MNRAS, 396, 1640
  
\bibitem[{{Ferland} {et~al.}(2013){Ferland}, {Porter}, {van Hoof}, {Williams},
  {Abel}, {Lykins}, {Shaw}, {Henney} \& {Stancil}}]{Ferland13}
{Ferland} G.~J., {et~al.}, 2013, Revista Mexicana de Astronomia y Astrofisica,
  49, 137

\bibitem[{{Gezari} {et~al.}(2006){Gezari}}]{GezariEtAl06}
{Gezari} S., {et~al.}, 2006, ApJ, 653, L25

\bibitem[{{Grier} {et~al.}(2013){Grier}, {Peterson}, {Horne}, {Bentz}, {Pogge},
  {Denney}, {De Rosa}, {Martini}, {Kochanek}, {Zu}, {Shappee}, {Siverd},
  {Beatty}, {Sergeev}, {Kaspi}, {Araya Salvo}, {Bird}, {Bord}, {Borman}, {Che},
  {Chen}, {Cohen}, {Dietrich}, {Doroshenko}, {Efimov}, {Free}, {Ginsburg},
  {Henderson}, {King}, {Mogren}, {Molina}, {Mosquera}, {Nazarov}, {Okhmat},
  {Pejcha}, {Rafter}, {Shields}, {Skowron}, {Szczygiel}, {Valluri} \& {van
  Saders}}]{GrierEtAl13}
{Grier} C.~J., {et~al.}, 2013, ApJ, 764, 47

\bibitem[{{Ivezic} {et~al.}(2008){Ivezic}, {Tyson}, {Acosta}, {Allsman},
  {Anderson}, {Andrew}, {Angel}, {Axelrod}, {Barr}, {Becker}, {Becla},
  {Beldica}, {Blandford}, {Bloom}, {Borne}, {Brandt}, {Brown}, {Bullock},
  {Burke}, {Chandrasekharan}, {Chesley}, {Claver}, {Connolly}, {Cook},
  {Cooray}, {Covey}, {Cribbs}, {Cutri}, {Daues}, {Delgado}, {Ferguson},
  {Gawiser}, {Geary}, {Gee}, {Geha}, {Gibson}, {Gilmore}, {Gressler}, {Hogan},
  {Huffer}, {Jacoby}, {Jain}, {Jernigan}, {Jones}, {Juric}, {Kahn}, {Kalirai},
  {Kantor}, {Kessler}, {Kirkby}, {Knox}, {Krabbendam}, {Krughoff}, {Kulkarni},
  {Lambert}, {Levine}, {Liang}, {Lim}, {Lupton}, {Marshall}, {Marshall}, {May},
  {Miller}, {Mills}, {Monet}, {Neill}, {Nordby}, {O'Connor}, {Oliver},
  {Olivier}, {Olsen}, {Owen}, {Peterson}, {Petry}, {Pierfederici},
  {Pietrowicz}, {Pike}, {Pinto}, {Plante}, {Radeka}, {Rasmussen}, {Ridgway},
  {Rosing}, {Saha}, {Schalk}, {Schindler}, {Schneider}, {Schumacher}, {Sebag},
  {Seppala}, {Shipsey}, {Silvestri}, {Smith}, {Smith}, {Strauss}, {Stubbs},
  {Sweeney}, {Szalay}, {Thaler}, {Vanden Berk}, {Walkowicz}, {Warner},
  {Willman}, {Wittman}, {Wolff}, {Wood-Vasey}, {Yoachim}, {Zhan} \& {for the
  LSST Collaboration}}]{IvezicEtAl08}
{Ivezic} Z., {et~al.}, 2008, ArXiv e-prints

\bibitem[{{Komossa}(2012)}]{Komossa2012}
{Komossa} S., 2012, in European Physical Journal Web of Conferences, Vol.~39, id.~02001

\bibitem[{{Komossa} \& {Bade}(1999)}]{KomossaBade99}
{Komossa} S., {Bade} N., 1999, A\&A, 343, 775

\bibitem[{{Komossa} {et~al.}(2003){Komossa}, {Burwitz}, {Hasinger}, {Predehl},
  {Kaastra} \& {Ikebe}}]{KomossaEtAl03}
{Komossa} S., {Burwitz} V., {Hasinger} G., {Predehl} P., {Kaastra} J.~S.,
  {Ikebe} Y., 2003, ApJ Lett., 582, L15

\bibitem[{{Komossa} {et~al.}(2008){Komossa}, {Zhou}, {Wang}, {Ajello}, {Ge},
  {Greiner}, {Lu}, {Salvato}, {Saxton}, {Shan}, {Xu} \&
  {Yuan}}]{KomossaEtAl08apjl}
{Komossa} S., {et~al.}, 2008, ApJ Lett., 678, L13

\bibitem[{{Liu} {et~al.}(2014){Liu}, {Li} \& {Komossa}}]{LiuEtAl14}
{Liu} F., {Li} S., {Komossa} S., 2014, ApJ Lett., 786, L103

\bibitem[{{Liu} \& {Chen}(2013)}]{LiuChen13}
{Liu} F.~K., {Chen} X., 2013, ApJ, 767, 18

\bibitem[{{MacFadyen} \& {Milosavljevi{\'c}}(2008)}]{MacFadyenMilosavljevic08}
{MacFadyen} A.~I., {Milosavljevi{\'c}} M., 2008, ApJ, 672, 83

\bibitem[{{Mayer} {et~al.}(2007){Mayer}, {Kazantzidis}, {Madau}, {Colpi},
  {Quinn} \& {Wadsley}}]{MayerEtAl07}
{Mayer} L., {Kazantzidis} S., {Madau} P., {Colpi} M., {Quinn} T., {Wadsley} J.,
  2007, Sci, 316, 1874

\bibitem[{{Murphy} {et~al.}(2013){Murphy}, {Chatterjee}, {Kaplan}, {Banyer},
  {Bell}, {Bignall}, {Bower}, {Cameron}, {Coward}, {Cordes}, {Croft}, {Curran},
  {Djorgovski}, {Farrell}, {Frail}, {Gaensler}, {Galloway}, {Gendre}, {Green},
  {Hancock}, {Johnston}, {Kamble}, {Law}, {Lazio}, {Lo}, {Macquart}, {Rea},
  {Rebbapragada}, {Reynolds}, {Ryder}, {Schmidt}, {Soria}, {Stairs}, {Tingay},
  {Torkelsson}, {Wagstaff}, {Walker}, {Wayth} \& {Williams}}]{MurphyEtAl12}
{Murphy} T., {et~al.}, 2013, Proc. Astron. Soc. Aust., 30, 6

\bibitem[{{Osterbrock} \& {Ferland}(2006)}]{OsterbrockFerland06}
{Osterbrock} D.~E., {Ferland} G.~J., 2006, {Astrophysics of gaseous nebulae and
  active galactic nuclei}

\bibitem[{{Pancoast} {et~al.}(2013){Pancoast}, {Brewer}, {Treu}, {Park},
  {Barth}, {Bentz} \& {Woo}}]{PancoastEtAl14}
{Pancoast} A., {Brewer} B.~J., {Treu} T., {Park} D., {Barth} A.~J., {Bentz}
  M.~C., {Woo} J.-H., 2013, ArXiv e-prints

\bibitem[{{Peterson}(1993)}]{Peterson93}
{Peterson} B.~M., 1993, PASP, 105, 247

\bibitem[{{Peterson}(2012)}]{Peterson12}
{Peterson} B.~M., 2012, Journal of Physics Conference Series, 372, 012008

\bibitem[{{Peterson} \& {Horne}(2004)}]{PetersonEtAl04}
{Peterson} B.~M., {Horne} K., 2004, Astronomische Nachrichten, 325, 248

\bibitem[{{Rees}(1988)}]{Rees88}
{Rees} M.~J., 1988, Nat, 333, 523

\bibitem[{{Rodriguez} {et~al.}(2006){Rodriguez}, {Taylor}, {Zavala}, {Peck},
  {Pollack} \& {Romani}}]{RodriguezEtAl06}
{Rodriguez} C., {Taylor} G.~B., {Zavala} R.~T., {Peck} A.~B., {Pollack} L.~K.,
  {Romani} R.~W., 2006, ApJ, 646, 49

\bibitem[{{Shen} \& {Loeb}(2010)}]{ShenLoeb10}
{Shen} Y., {Loeb} A., 2010, ApJ, 725, 249

\bibitem[{{Sillanp{\"a}{\"a}} {et~al.}(1988){Sillanp{\"a}{\"a}}, {Haarala},
  {Valtonen}, {Sundelius} \& {Byrd}}]{SillanpaaEtAl88}
{Sillanp{\"a}{\"a}} A., {Haarala} S., {Valtonen} M.~J., {Sundelius} B., {Byrd}
  G.~G., 1988, ApJ, 325, 628

\bibitem[{{Snedden} \& {Gaskell}(1999)}]{SneddenGaskell99}
{Snedden} S.~A., {Gaskell} C.~M., 1999, ApJ Lett., 521, L91

\bibitem[{{Strubbe} \& {Quataert}(2009)}]{StrubbeQuataert09}
{Strubbe} L.~E., {Quataert} E., 2009, MNRAS, 400, 2070

\bibitem[{{Tanaka}(2013)}]{Tanaka13}
{Tanaka} T.~L., 2013, MNRAS, 434, 2275

\bibitem[{{Ulmer}(1999)}]{Ulmer99}
{Ulmer} A., 1999, ApJ, 514, 180

\bibitem[{{Valtonen} {et~al.}(2008){Valtonen}, {Lehto}, {Nilsson}, {Heidt},
  {Takalo}, {Sillanp{\"a}{\"a}}, {Villforth}, {Kidger}, {Poyner}, {Pursimo},
  {Zola}, {Wu}, {Zhou}, {Sadakane}, {Drozdz}, {Koziel}, {Marchev}, {Ogloza},
  {Porowski}, {Siwak}, {Stachowski}, {Winiarski}, {Hentunen}, {Nissinen},
  {Liakos} \& {Dogru}}]{ValtonenEtAl08}
{Valtonen} M.~J., {et~al.}, 2008, Nat, 452, 851

\bibitem[{{Yuan} {et~al.}(2013)}]{YuanEtAl13}
Yuan W., {et~al.}, 2013, {NAOC internal document}

\bibitem[{{Zauderer} {et~al.}(2011){Zauderer}, {Berger}, {Soderberg}, {Loeb},
  {Narayan}, {Frail}, {Petitpas}, {Brunthaler}, {Chornock}, {Carpenter},
  {Pooley}, {Mooley}, {Kulkarni}, {Margutti}, {Fox}, {Nakar}, {Patel},
  {Volgenau}, {Culverhouse}, {Bietenholz}, {Rupen}, {Max-Moerbeck}, {Readhead},
  {Richards}, {Shepherd}, {Storm} \& {Hull}}]{ZaudererEtAl11}
{Zauderer} B.~A., {et~al.}, 2011, Nat, 476, 425

\end{thebibliography}

\end{document}